\documentclass[a4paper]{article}

\usepackage{INTERSPEECH2021, url}
\title{Unsupervised Multi-Target Domain Adaptation for Acoustic Scene Classification}
\name{Dongchao Yang$^1$, Helin Wang$^1$, Yuexian Zou$^{1,2}$}
\address{
  $^1$ADSPLAB, School of ECE, Peking University, Shenzhen, China\\
  $^2$Peng Cheng Laboratory, Shenzhen, China}
\email{dongchao98@stu.pku.edu.cn, \{wanghl15,zouyx\}@pku.edu.cn}

\begin{document}

\maketitle
\begin{abstract}
It is well known that the mismatch between training (source) and test (target) data distribution will significantly decrease the performance of acoustic scene classification (ASC) systems. To address this issue, domain adaptation (DA) is one solution and many unsupervised DA methods have been proposed. These methods focus on a scenario of single source domain to single target domain. However, we will face such problem that test data comes from multiple target domains. This problem can be addressed by producing one model per target domain, but this solution is too costly. In this paper, we propose a novel unsupervised multi-target domain adaption (MTDA) method for ASC, which can adapt to multiple target domains simultaneously and make use of the underlying relation among multiple domains. Specifically, our approach combines traditional adversarial adaptation with two novel discriminator tasks that learns a common subspace shared by all domains. Furthermore, we propose to divide the target domain into the easy-to-adapt and hard-to-adapt domain, which enables the system to pay more attention to hard-to-adapt domain in training. The experimental results on the DCASE 2020 Task 1-A dataset and the DCASE 2019 Task 1-B dataset show that our proposed method significantly outperforms the previous unsupervised DA methods. 
\end{abstract}
\section{Introduction}
\begin{figure}[t]
  \centering
  \includegraphics[width=\linewidth]{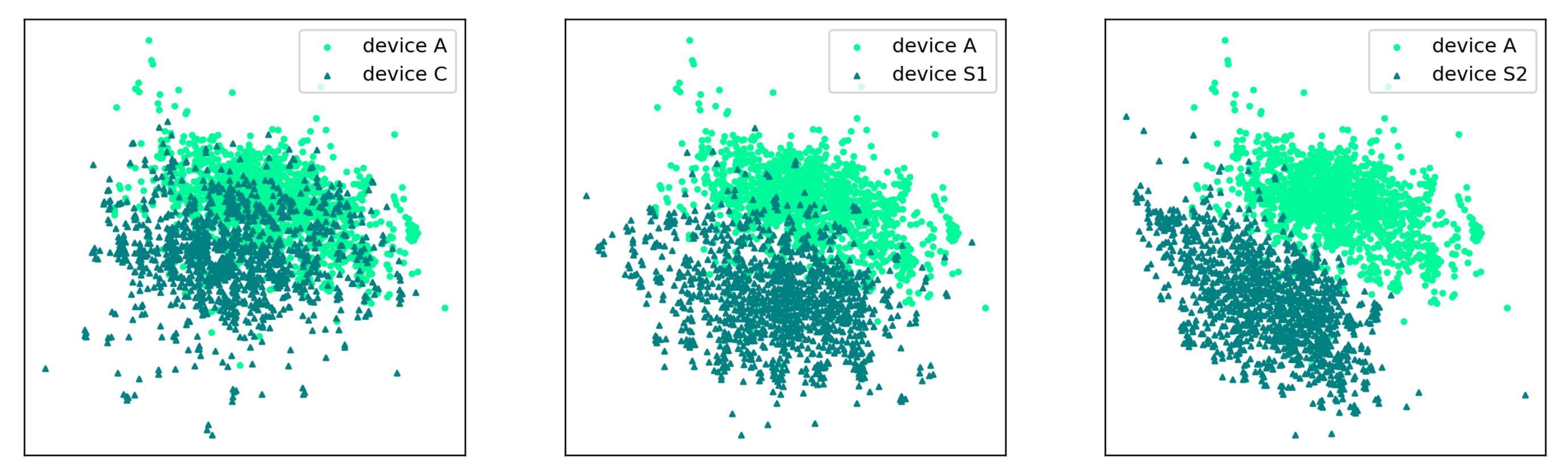}
  \caption{Visualization of mel-spectrum of audio data from DCASE2020 dataset \cite{heittola2020acoustic} by t-SNE. We choose small subset of samples from device A, and parallel recordings from three lower quality devices (devices C, S1 and S2).}
  \label{fig:distribution}
  \vspace*{-\baselineskip}
\end{figure}

Acoustic scene classification (ASC) is the task of assigning a scene label (e.g., “Tram”, “Park”) to an audio recording. ASC has recently been tackled with many deep learning methods \cite{hershey2017cnn,piczak2015environmental,dorfer2018acoustic,wang2020environmental,kong2020panns}. However, many ASC systems tend to be susceptible to the effects of domain shift, when training and test audio are recorded by different devices. Figure \ref{fig:distribution} shows that different recording devices lead to the change of data distribution. To address the problem of mismatched recording devices, many methods have been proposed, such as data augmentation \cite{nguyen2019acoustic,hu2020two}, spectrum correction \cite{kosmider2020spectrum,nguyen2020acoustic} and domain adaptation (DA) \cite{primus2019exploiting}. Although these methods got good performance, they trained with both labeled source- and target-domain samples. In this paper, we investigate the unsupervised domain adaptation (UDA) scenario, i.e., the acoustic scene labels of the target domain are not known during the adaptation part.
 
Many UDA methods \cite{ben2010theory,bousmalis2016domain,wang2020continuously} have been proposed in computer vision field, but only a few studies (such as \cite{gharib2018unsupervised,drossos2019unsupervised,takeyama2020robust,wang2019domain,mezza2021unsupervised}) have applied  UDA techniques to ASC models. In \cite{wang2019domain}, authors follow a unsupervised domain adaptation neural network \cite{ganin2015unsupervised}, and introduces it to learn a common subspace for the ASC problem. In \cite{takeyama2020robust}, authors follow maximum classifier discrepancy \cite{saito2018maximum}, which can properly consider distributions of each class within domains.

\begin{table}[h]\footnotesize
\caption{Comparison of accuracy on DCASE2020 dataset \cite{heittola2020acoustic}. We use device A as source domain, devices B, C, S1 and S2 as target domains. DANN means that combining devices B, C, S1 and S2 as a single domain, then training once. DANN-respective means that we train (A $\rightarrow$ B), (A $\rightarrow$ C), (A $\rightarrow$ S1), (A $\rightarrow$ S2) separately.}
\centering
\begin{tabular}{|l|l|l|l|l|l|}
\hline
model           & \begin{tabular}[c]{@{}l@{}}B(\%)\end{tabular} & \begin{tabular}[c]{@{}l@{}}C(\%)\end{tabular} & \begin{tabular}[c]{@{}l@{}}S1(\%)\end{tabular} & \begin{tabular}[c]{@{}l@{}}S2(\%)\end{tabular}  \\ \hline
DANN \cite{wang2019domain}            & 47                                             & 53.3                                           & 35.8                                            & 28.5                                                                                        \\ \hline
DANN-respective & 48.2                                           & 53.9                                           & 40.3                                            & 34.8                                                                                        \\ \hline
\end{tabular}
\label{tab:my-table}
\vspace{-0.5cm}
\end{table}

But all of these methods focus on pairwise adaptation settings from single source domain to single target domain. However,
in reality, test data may come from multiple target domains, e.g., audio data is recorded by multiple devices. When test data consist of multiple target domains, there are two common solutions. One is
combining all target domains as a single target domain \cite{gharib2018unsupervised,drossos2019unsupervised,takeyama2020robust,wang2019domain}, and then applying once pairwise adaptation. The other is applying pairwise adaptation for each target domain separately. As Table \ref{tab:my-table} shows, experimental results indicate that combining all target domains as a single one will decrease performance, and in Section \ref{sec:previous approach} we also give theoretical demonstration. Although we can produce one model per target domain, this approach becomes costly and impractical in applications with a growing number of target domains. In addition, applying pairwise adaptation approach may be suboptimal, as it ignore the underlying relation among multiple domains.

In this paper, we propose a novel unsupervised multi-target domain adaption (MTDA) method for ASC, which can adapt to multiple target domains simultaneously and make use of the underlying relation among multiple domains. Our approach is based on two technical insights. The first technical insight is to learn a common subspace shared by both the source and target domains, which enables all domains to have same data distribution in the feature space. Specifically, inspired by GAN \cite{goodfellow2014generative} and DANN \cite{wang2019domain}, we make use of the adversarial relationship between modules Feature (F) and Discriminator (D) to learn domain-invariant feature in the feature space. Unlike previous methods \cite{gharib2018unsupervised,wang2019domain} which apply D in the binary classification task, we utilize D in the multi-classification or regression task. The second technical insight is to divide target domain into easy-to-adapt or hard-to-adapt domain. Intuitively, if the target domain is close enough to the source domain, the feature extracted from itself tends to result in accurate classification. Furthermore, we note that different devices cause different degrees of domain shift, and the target domain with severe domain shift has poor performance and is hard to adapt. Consequently, we should pay more attention to these hard-to-adapt domains during the training process.
Our contributions are as follows:

(1) We theoretically demonstrate why the pairwise adaptation methods cannot performance well when combining all target domains as a single one.

(2) We propose to divide the target domain into the easy-to-adapt or the hard-to-adapt domain.

(3) We propose the first unsupervised multi-target domain adaptation approach for ASC, which improves the performance of ASC unsupervised DA over the previous SOTA methods.
\begin{figure}[t]
  \centering
  \includegraphics[width=0.6\linewidth]{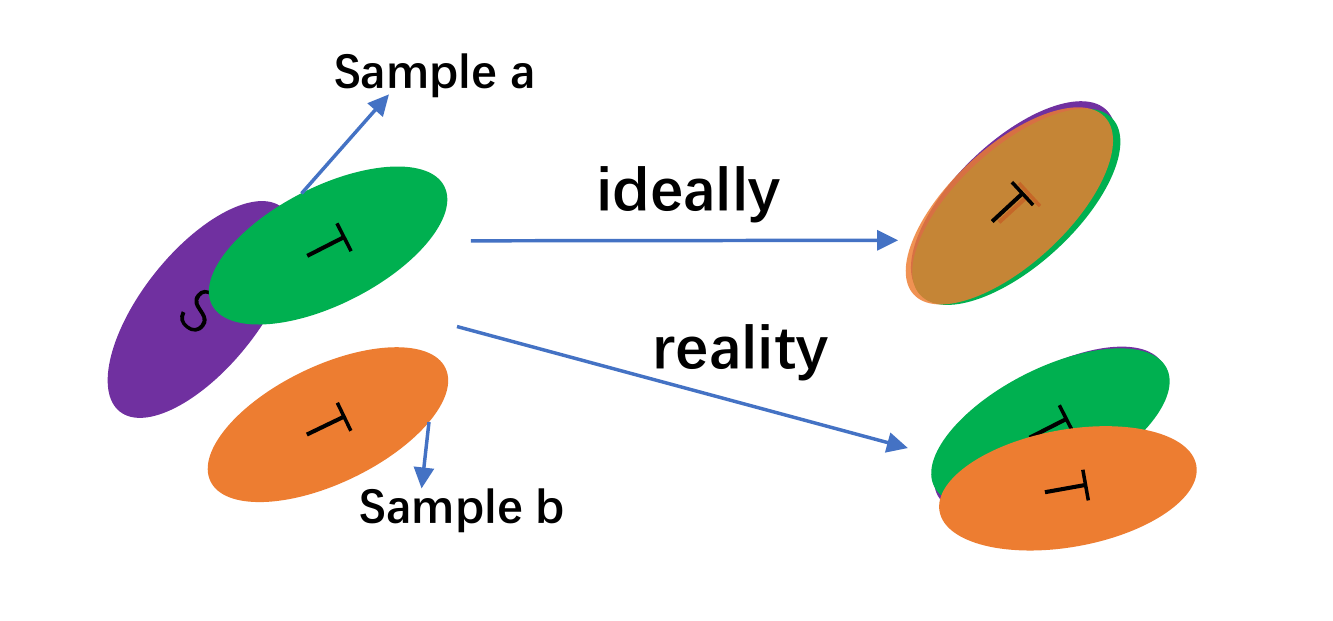}
  \caption{Ideally, we want to align all domains. In fact, hard-to-adapt domain may not be aligned well.}
  \label{fig:adapt-pic}
  \vspace*{-\baselineskip}
\end{figure}

\section{Vanilla method} \label{sec:previous approach}
In this section, we will theoretically demonstrate why the pairwise adaptation methods cannot performance well when combining different devices data as a single one. The following proof takes DANN \cite{wang2019domain} as an example.

In DANN \cite{wang2019domain}, authors propose to learn an encoder $E$ and a predictor $C$ such that the distribution of the encodings $z=E(\vec{x})$ (where \vec{x} denotes mel-spectrum of audio) from target and source domains are aligned so that all scene labels can be accurately predicted by the shared predictor $C$. This is achieved by the adversarial relationship between $E$ and discriminator $D$. $D$ predicts  $\vec{z}$ coming from source or target domain. The domain label is defined as $\vec{d}=[0,1]^T$ (which stands for source domain) or $\vec{d}=[1,0]^T$ (which stands for target domain).
\begin{equation}\label{}
    L_d=||D(\vec{z})-\vec{d}||^2_{2}
\end{equation}
\begin{equation}\label{}
    L_e=L_y(C(\vec{z}),\vec{y})-\lambda_d*L_d
\end{equation}
Where $L_d$ denotes the domain classification loss of discriminator, $L_y$ denotes the scene classification loss of predictor, and $y$ denotes the scene label of audio. $L_e$ denotes the training objective function, which minimizes the scene classification loss and meanwhile maximizes the domain classification loss. The parameter $\lambda_d$ controls the trade-off between $L_y$ and $L_d$. In order to simplify our prove, we only consider the relationship between $D$ and $E$. Formally, DANN performs a maximin optimization with the value function $V_d(E,D)$.
\begin{equation}\label{}
    \max \limits_{E} \min \limits_{D} V_d(E,D)=\mathbb{E}(L_d(D(E(\vec{x})), \vec{d}))
\end{equation}
Where $\mathbb{E}$ denotes math expectation. When E is fixed, the optimal D is shown as formula (\ref{formula 4}).
\begin{equation}\label{formula 4}
\begin{split}
    D^*_{E}=\mathop{\arg\min}_{D} \mathbb{E}_{(\vec{z},\vec{d})\sim p(\vec{z},\vec{d})}[||D(\vec{z})-\vec{d})||^2_{2}]\\
    =\mathop{\arg\min}_{D} \mathbb{E}_{\vec{z}\sim p(\vec{z})}\mathbb{E}_{\vec{d}\sim p(\vec{d}|\vec{z})}[||D(\vec{z})-\vec{d})||^2_{2}]
    \end{split}
\end{equation}
Formula (\ref{formula 4}) is equivalent to minimize $\mathbb{E}_{\vec{d}\sim p(\vec{d}|\vec{z})}[||D(\vec{z})-\vec{d}||^2_{2}]$.
\begin{equation}\label{formula 5}
\begin{split}
    \mathbb{E}_{\vec{d}\sim p(\vec{d}|\vec{z})}[||D(\vec{z})-\vec{d}||^2_{2}] =
    \mathbb{E}_{\vec{d}\sim p(\vec{d}|\vec{z})}[\vec{d^2}] - \\ -2D(\vec{z})\mathbb{E}_{\vec{d}\sim p(\vec{d}|\vec{z})}[\vec{d}] +D(\vec{z})^2
    \end{split}
\end{equation}
Formula (\ref{formula 5}) is a quadratic form of $D(\vec{z})$  which achieves the minimum at $D(\vec{z})=\mathbb{E}_{\vec{d}\sim p(\vec{d}|\vec{z})}[\vec{d}]$.\\
Assuming that $D$ always achieves its optimum, the maximin optimization in formula (3) can be reformulated as maximizing $C_d(E)$ where
\begin{equation}\label{}
    C_d(E) \triangleq \min \limits_{D} V_d(E,D)=V_d(E,D^*_{E})
\end{equation}
When $D$ achieves its optimal, we fix it and then update $E$ to maximize $C_d(E)$. Considering such scenario, there are one source domain and two target domains. As Figure \ref{fig:adapt-pic} shows, two \textit{samples} (a and b) are on different target domains, and \textit{sample a} is easier to adapt because it is closer to source domain. But when updating parameters of $E$ (denoted as $\theta$) by gradient descent, we find that two samples have the same gradient $ \frac{\partial L_d}{\partial \theta} $, because $\mathbb{E}_{\vec{d} \sim p(\vec{d}|\vec{z})}[\vec{d}]$ is a constant and they have the same domain label $\vec{d}$. In fact, a target domain is far from source domain, which needs a greater gradient. So when combining all target domain as a single one, some difficult-adapt domains cannot be adapted well.
\begin{figure}[t]
  \centering
  \includegraphics[width=0.9\linewidth, height=0.35\linewidth]{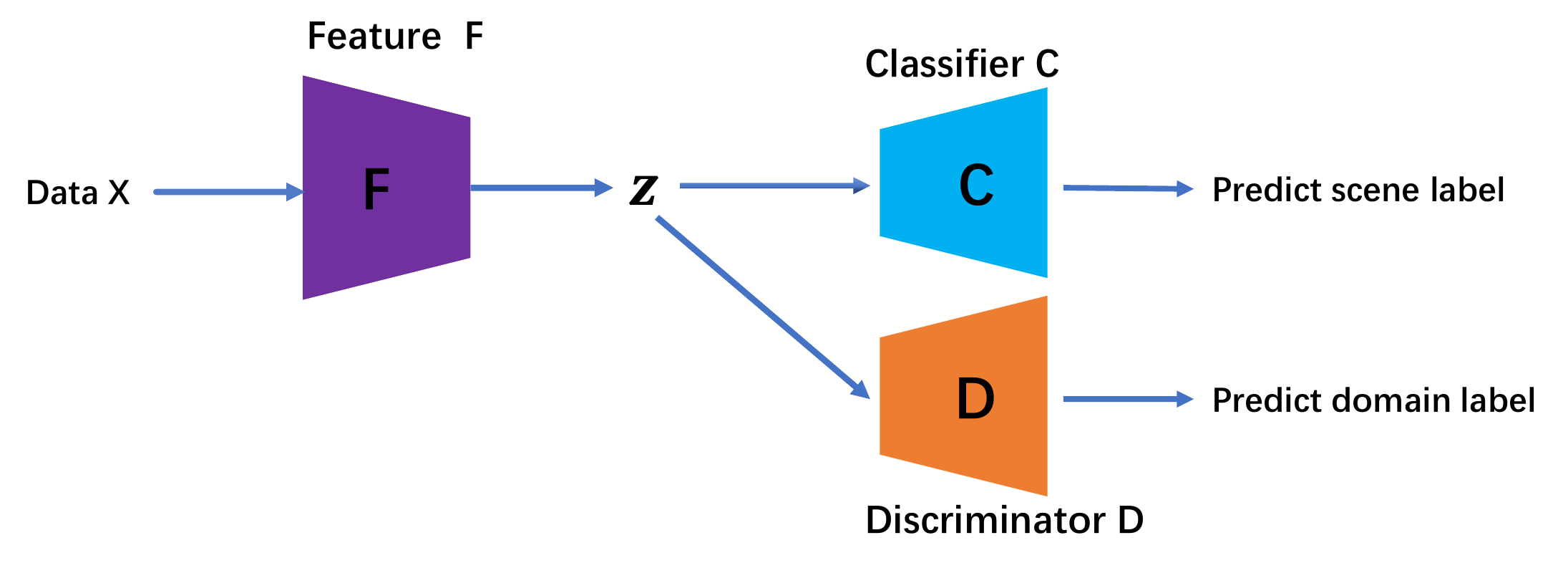}
  \caption{Multi-target domain adaptation network.}
  \label{fig:model-pic}
  \vspace*{-\baselineskip}
\end{figure}
\section{Proposed method} \label{sec:proposed-method}
In this section, we formalize the problem of adaptation among multi-target domain, and describe our methods for addressing the problem. Specifically, we will introduce domain distance and domain index in Section \ref{subsec: distance}, then we propose two novel discriminator tasks base on domain index in Section \ref{subsuc: multi-classification} and Section \ref{subsec: regression problem}.
\subsection{Problem formulation}
We consider a multi-class (K-class) classification problem for ASC. Let $(X,Y,D)=\{(\vec{x}_i,\vec{y}_i,\vec{d}_i)\}^N_{i=0}$ be a collection of $M$ domains (a labeled source domain and $M-1$ unlabeled target domains). $\vec{x}_i$ denotes the i-th audio sample. $\vec{y}_i$ denotes scene label of the i-th audio, and $\vec{y}_i=[y^0_i,y^1_i,y^2_i,...,y^K_i]$. $\vec{d}_i$ denotes domain label of the i-th audio, and $\vec{d}_i=[d^0_i,d^1_i,d^2_i,...,d^{M-1}_i]$. Scene labels $\vec{y}_i$ and domain label $\vec{d}_i$ are both one-hot vectors. $\vec{y}_i$ is only available for the source samples, but $\vec{d}_i$ is available for all samples, because domain label is generated by ourselves.
Figure \ref{fig:model-pic} shows the diagram of the proposed method. For any $\vec{x}_i$, Feature (F) is used to extract its features $\vec{z}_i$, and $\vec{z}_i=F(\vec{x}_i)$. The Classifier (C) tries to predict scene label, and the Discriminator (D) tries to predict domain label. F aims to project acoustic features from different domains into one subspace where the features are scene-discriminative and domain-invariant. D tries to discriminate which domains the input audio recording comes from.
\subsection{Domain distance and domain index} \label{subsec: distance}
\textbf{Domain distance} We can find two useful things by analysing Figure \ref{fig:distribution} and Table \ref{tab:my-table}. Firstly, different devices cause different degrees of domain shift. Secondly, the greater the domain shift of the target domain, the poorer performance the target domain can obtain and the harder to adapt, e.g., device S1 and S2. Based on these two facts, we introduce domain distance to describe the extent of target domain shift. Domain distance is defined as the distance between the target domain and the source domain. To calculate the domain distance, the parallel datas (it means these audios are recorded in the same environment simultaneously, but using different devices) are used. We assume that parallel samples contain the same information about the acoustic scenes and differ only due to device characteristics. To calculate domain distance conveniently, we reduce the dimension of the mel-spectrum of audio data (\vec{x}) by t-SNE algorithm \cite{van2008visualizing}. We define $\vec{a}$ as the reduced dimension data, and $\vec{a}=tSNE(\vec{x})$. Then we use formula (\ref{formula distance-i}) to get domain distance.
\begin{equation}\label{formula distance-i}
\setlength{\abovedisplayskip}{3pt}
\setlength{\belowdisplayskip}{3pt}
    Distance_i = \frac{1}{N}\sum_{j=0}^{N-1} ||\vec{a}_{i,j}-\vec{a}^*_{i,j}||
\end{equation}
Where $Distance_i$ denotes the distance from the i-th target domain to the source domain. $\vec{a}_{i,j}$ denotes the j-th data of the i-th target domain, and $\vec{a}^*_{i,j}$ is the data of source domain parallel to $\vec{a}_{i,j}$. N denotes the number of parallel data.
We rank target domains according to their domain distance. If the target domain has high ranking, it is hard to adapt.\\
\textbf{Domain index} Domain index is used to quantify domain distance, which indicates the relative distance between the target and the source domain. So the ranking of target domain is used as their domain index. The index of source domain is set as 0.
\subsection{Multi-classification task} \label{subsuc: multi-classification}
According to previous methods \cite{gharib2018unsupervised,drossos2019unsupervised,wang2019domain}, it is easy to consider letting discriminator to do a multi-classification task.  
The feed-forward process are summarized as formula (\ref{formula feed-forward}) shows.
\begin{equation}\label{formula feed-forward}
\setlength{\abovedisplayskip}{4pt}
\setlength{\belowdisplayskip}{4pt}
    \vec{z}=F(\vec{x}), \tilde{\vec{y}}=C(\vec{z}), \tilde{\vec{d}}=D(\vec{z})
\end{equation}

The update process is using backpropagation algorithm. For Classifier and Discriminator, loss function is cross-entropy function. For Feature, loss function is defined as formula (\ref{formula 9}) shows, where $L_y$ and $L_d$ denote the error of scene classification and domain classification, respectively. $\lambda_d$ is a hyper-parameter.
\begin{equation}\label{formula 9}
\setlength{\abovedisplayskip}{2pt}
\setlength{\belowdisplayskip}{2pt}
L_{f} = L_y(C(\vec{z}),\vec{y})-\lambda_d*L_d
\end{equation}

Inspired by focal loss \cite{lin2017focal}, we modify cross-entropy function of Discriminator according to domain index. Our motivation is paying more attention to the hard-to-adapt domain. The improved loss function is defined as formula (\ref{improved cross-function}) shows. Where $u_i$ denotes the domain index of i-th sample, $N$ denotes the number of data. $\vec{d}_i$ and $\vec{\tilde{d}}_i$ denote the domain label and the predicted result of i-th sample, respectively. $T$ is a hyper-parameter, in our experiments, we set $T=10$.

\begin{equation}\label{improved cross-function}
\setlength{\abovedisplayskip}{1pt}
\setlength{\belowdisplayskip}{1pt}
L_{d} = -\sum^{N-1}_{i=0}\frac{u_i+1}{T}\vec{d}_{i}\log\tilde{\vec{d}}_{i}
\end{equation}
\subsection{Regression task} \label{subsec: regression problem}
We note that the domain index plays the role of a distance metric, i.e., it captures a similarity distance between the target domain and the source domain. So the domain index also can be viewed as domain label, and we consider letting discriminator to regress the domain index using a distance-base loss, such as $L_2$ loss. The feed-forward process and the update process are similar with multi-classification task in Section \ref{subsuc: multi-classification}, the difference is that Discriminator just needs to predict domain index (denoted as $u$), and the loss function of Discriminator is defined as $L_d=||D(\vec{z})-u||^2_2$.\\
\section{Experiment} \label{sec:experiment}
\subsection{Datasets and metrics}
\textbf{Datasets} The DCASE 2019 task1B dataset \cite{mesaros2019acoustic} and the DCASE 2020 task1A dataset \cite{heittola2020acoustic} contain 10s segments, recorded at 48kHz and spanning 10 classes.\\
\textbf{Evaluation metrics}  For all the experiments, we use the accuracy of classification as the evaluation metric, which is one of the most commonly used metrics for audio classification \cite{virtanen2018computational}.
\subsection{Experiments on DCASE2019 dataset}
Our first experiment evaluates on DCASE 2019 task1B \cite{mesaros2019acoustic}. In the DCASE2019 task1B dataset, all data are recorded by device A, B and C. The data of device A is regarded as source domain and that of device B and C as two target domains.\\
\textbf{Experimental setups} To make a fair comparison, we use the same model structure and experimental
setting as DANN \cite{wang2019domain}. See \cite{wang2019domain} for more details about DANN.\\
\textbf{Experimental results and analysis} Table \ref{tab:dcase2019-results} demonstrates the performance of our proposed methods and DANN on DCASE 2019 task1B dataset. We set three different experiments for MTDA. MTDA-R denotes that we make Discriminator do regression task. MTDA-C1 denotes that we make Discriminator do multi-classification task and choose cross-entropy as loss function. MTDA-C2 denotes loss function is improved cross-entropy according to formula (\ref{improved cross-function}). All of our three experiments perform better than DANN, which confirms
the effectiveness of MTDA. MTDA-C2 performs better than MTDA-C1, which shows dividing target domain into easy-to-adapt and hard-to-adapt domain is very useful.
\begin{table}[t]\footnotesize
\caption{Comparison of accuracy on DCASE 2019 task1b dataset.}
\label{tab:dcase2019-results}
\centering
\begin{tabular}{|l|l|l|l|l|}
\hline
model               & A(\%)         & B(\%)         & C(\%)         & B\&C(\%)      \\ \hline
DANN \cite{wang2019domain}           & 60.3          & 42.7          & 46.7          & 44.7          \\ \hline
\textbf{MTDA-R(ours)}  & \textbf{63.9}          & 40.6          & 52.1          & 46.4         \\ \hline
\textbf{MTDA-C1(ours)} & 62.4 & 42.4 & 50.3          & 46.3         \\ \hline
\textbf{MTDA-C2(ours)} & 63     &    \textbf{45.5}          & \textbf{54.4} & \textbf{49.9} \\ \hline
\end{tabular}
\vspace{-0.5cm}
\end{table}
\subsection{Experiments on DCASE2020 dataset}
\begin{figure*}[t]
  \centering
  \includegraphics[width=0.95\linewidth,height=0.2\linewidth]{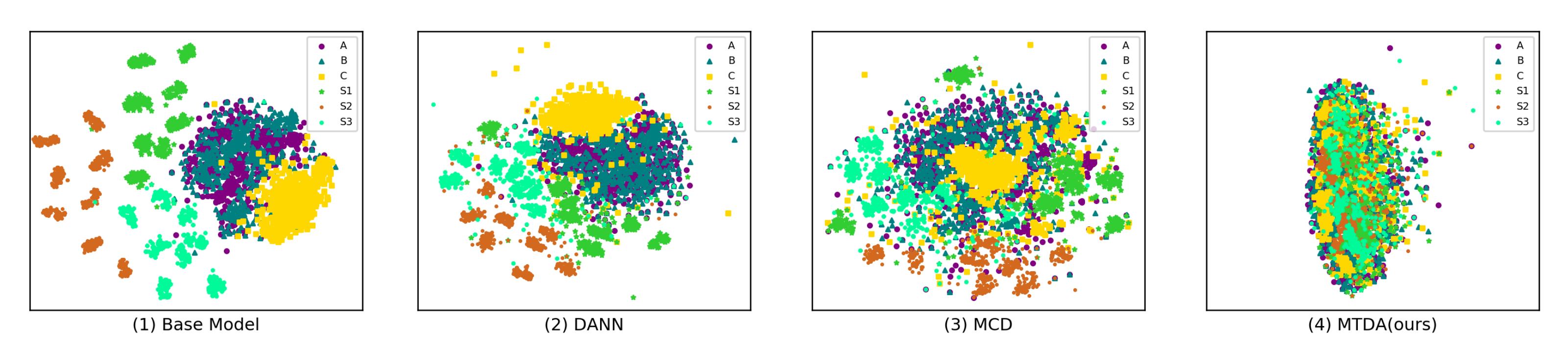}
  \caption{Visualization of feature ($\vec{z}$) by t-SNE, which is extracted from audio data (\vec{x}) by module Feature.}
  \label{fig:compare-model}
  \vspace*{-\baselineskip}
\end{figure*}
Our second experiment evaluates on DCASE 2020 task1A \cite{heittola2020acoustic}. In our experiments, we take the data of device A as source domain and that of device B, C, S1, S2, S3 as target domains. Furthermore, we set the data of device S4, S5 and S6 as unseen domains, these data only available on test stage. For device A, 10215 segments audio are used to train.
For device B, C, S1-S3, 750 segments audio are used to adapt, these audio are regarded as unlabeled data. In the test set, 330 segments audio for each device are used.\\
\textbf{Previous methods}
We compare our method with previous state-of-the-art unsupervised DA methods for ASC including DANN \cite{wang2019domain}, UADA \cite{gharib2018unsupervised}, W-UADA \cite{drossos2019unsupervised}, MCD \cite{takeyama2020robust}, MMD \cite{long2015learning}. To fairly compare with these methods, we choose two baseline models. One is DCASE model \cite{kong2019cross}, which consists of 8 layers CNN. The other is Resnet14 model \cite{he2016deep}. All approaches are implemented based on these two baselines. We have released all the code at \url{https://github.com/yangdongchao/interspeech2021_MTDA}. \\
\textbf{Preprocessing} All the raw audios are resampled to 32kHz and fixed to a certain length of 10s. The short time Fourier transform
(STFT) is then applied on the audio signals to calculate spectrograms, with a window size of 32ms and a hop size of 15.6ms. 64 mel filter banks are applied on the spectrograms followed by a
logarithmic operation to extract the log mel spectrograms.\\
\textbf{Experiment setting}
In the training phase, for UADA and W-UADA, the RMSProp \cite{tieleman2012lecture} optimizer is used. For other methods, the Adam algorithm \cite{kingma2014adam} is employed as the optimizer. All models are trained with an initial learning rate of 0.002. The batch size is set to 32 and training epoch is 200. $\lambda_d$ is chosen from \{0.2, 0.5, 1.0, 2.0, 5.0, 8.0, 10.0\}. In our experiments, we never use any data augmentation methods.\\
\textbf{Experimental results and analysis}
Table \ref{tab:dcase2020-cnn} and Table \ref{tab:dcase2020-resnet} demonstrate the performance of our proposed MTDA and other state-of-the-art methods. For base model \cite{kong2019cross, he2016deep}, we only train models on source data and do not use any domain adaptation. Due to the difference in the acquisition equipment greatly affects the characteristics of the audio signals, test data without domain adaptation is difficult to obtain good performance on the classifier trained in the source domain. DANN-respective denotes that we apply DANN for each target domain separately. For methods \cite{long2015learning,gharib2018unsupervised,drossos2019unsupervised,wang2019domain,takeyama2020robust}, we combine the data of device B, C, S1-S3 as one target domain.
Although DANN and MCD can get good performance on easy-to-adapt domain, such as device B and C, they cannot perform well on hard-to-adapt domain, such as device S1-S3. On the contrary, our methods can get better performance on device S1-S3. Our methods perform better than DANN-respective, which shows MTDA can make better use of the underlying relation among multiple domains. Furthermore, our methods get better performance on unseen domains (device S4-S6), which shows our methods have good generalization. We note that previous methods \cite{long2015learning,gharib2018unsupervised,drossos2019unsupervised,takeyama2020robust} cannot successfully classify source samples (device A) compared with baseline. It means that the loss of source domain information is influenced by adaptation process. Although our methods cannot overcome this problem completely, our methods also yield comparable results with baseline.\\
\textbf{Comparison between MTDA-C1 and MTDA-C2}
MTDA-C2 achieves higher accuracy than MTDA-C1 for the reason that MTDA-C2 pays more attention to hard-to-adapt domain.\\
\textbf{Comparison between MTDA-C and MTDA-R}
MTDA-R also gets good performance, which validates the effectiveness of making Discriminator regress domain index. MTDA-R can get comparable results with MTDA-C2 for the reason that domain index plays the role of a distance metric, which means that regressing domain index by the Discriminator can also benefit the hard-to-adapt domain. Furthermore, MTDA-R achieves higher accuracy on unseen domain than MTDA-C.\\
\textbf{Feature visualization}
T-SNE \cite{van2008visualizing} is used to visualize adaptation results for different methods. Figure \ref{fig:compare-model} (a) shows the results of base model \cite{kong2019cross}. Figure \ref{fig:compare-model} (b) and Figure \ref{fig:compare-model} (c) show adaptation results of DANN and MCD model, and they can adapt to easy-to-adapt domains, such as device B and C, but they cannot adapt to hard-to-adapt domains well, such as device S1-S3. Figure \ref{fig:compare-model} (d) shows adaptation results of MTDA-C2. We can find that almost all distribution of target domains align with the source domain.
\begin{table}[t]\footnotesize
\centering
\caption{Accuracy (\%) comparison of different methods on DCASE 2020 task1a development set. The base model is DCASE model, and all methods are based on this model.}
\label{tab:dcase2020-cnn}
\begin{tabular}{|l|c|c|c|c|}
\hline
model                  & \multicolumn{1}{l|}{A} & \multicolumn{1}{l|}{B\&C} & \multicolumn{1}{l|}{S1-S3} & \multicolumn{1}{l|}{S4-S6} \\ \hline
Base(DCASE) \cite{kong2019cross}                   & \textbf{68.8}                   & 41.1                      & 23.6                       & 24.9                       \\ \hline
MMD \cite{long2015learning}                    & 66.1                   & 42.6                      & 25.7                       & 30.8                       \\ \hline
UADA \cite{gharib2018unsupervised}                   & 59.4                   & 44.7                      & 37.4                       & 32.8                       \\ \hline
W-UADA \cite{drossos2019unsupervised}                & 67.0                     & 46.4                      & 30.9                       & 30.9                       \\ \hline
DANN \cite{wang2019domain}                   & 68.2                   & 51.5                      & 37.1                       & 36.7                       \\ \hline
DANN-respective       & 68.2                   & 54.1             & 37.8                       & -                          \\ \hline
MCD \cite{takeyama2020robust}                    & 63.9                   & 52.1                      & 33.0                         & 36.8                       \\ \hline
\textbf{MTDA-R(ours)}  & 68.5          & 52.6                      & \textbf{41.7}              & \textbf{41.5}              \\ \hline
\textbf{MTDA-C1(ours)} & 67.9                   & 52.0                        & 39.3                       & 37.7                       \\ \hline \textbf{MTDA-C2(ours)}  & 68.2
& \textbf{54.5}             & \textbf{41.7}              & 41.4                       \\ \hline
\end{tabular}
\end{table}
\begin{table}[t]\footnotesize
\centering
\caption{Accuracy (\%) comparison of different methods on DCASE 2020 task1a development set. The base model is Resnet model, and all methods are based on this model.}
\label{tab:dcase2020-resnet}
\begin{tabular}{|l|c|c|c|c|}
\hline
model                  & \multicolumn{1}{l|}{A} & \multicolumn{1}{l|}{B\&C} & \multicolumn{1}{l|}{S1-S3} & \multicolumn{1}{l|}{S4-S6} \\ \hline
Base(Resnet14) \cite{he2016deep}           & \textbf{67.6}                   & 39.8                      & 22.4                       & 25.6                                  \\ \hline
MMD \cite{long2015learning}                    & 62.7                   & 45.9                     & 26.8                       & 29.3                                  \\ \hline
UADA \cite{gharib2018unsupervised}                  & 63.3                   & 47.3                      & 37.6                       & 39.6                                  \\ \hline
W-UADA \cite{drossos2019unsupervised}                & 64.2                   & 45.0                        & 33.2                       & 34.3                                  \\ \hline
DANN \cite{wang2019domain}                   & 65.9                   & 50.2                      & 35.1                       & 37.9                                  \\ \hline
DANN-respective        & 65.3                   & 51.1                      & 38.6                       & -                                     \\ \hline
MCD \cite{takeyama2020robust}                   & 65.2                   & 51.2             & 30.2                       & 33.1                                  \\ \hline
\textbf{MTDA-R(ours)}  & 67.2                   & 51.5                        & 41.9              & \textbf{39.9}                         \\ \hline
\textbf{MTDA-C1(ours)} & 65.5                   & 50.5                      & 37.6                       & 39.1                                  \\ \hline
\textbf{MTDA-C2(ours)} & 67.5          & \textbf{53.2}                      & \textbf{42.3}                       & 39.4                                  \\ \hline
\end{tabular}
\vspace*{-\baselineskip}
\end{table}
\section{Conclusions}
A novel multi-target domain adaption (MTDA) method for ASC has been proposed. Our method can adapt multiple target domains simultaneously and make use of the underlying relation among domains. Experimental results on the ASC tasks and visualization analysis validate the advantages of our method.

\bibliographystyle{IEEEtran}

\bibliography{mybib}


\end{document}